\def\ra{\rangle}
\def\la{\langle}
\def\be{\begin{equation}}
\def\ee{\end{equation}}
\def\ba{\begin{array}}
\def\ea{\end{array}}

\def\Cb{{\Bbb C}}

\documentclass[pra,showpacs,twocolumn,amsmath]{revtex4}
\usepackage{epsfig}
\def\qed{\leavevmode\unskip\penalty9999 \hbox{}\nobreak\hfill
     \quad\hbox{\leavevmode  \hbox to.77778em{%
               \hfil\vrule   \vbox to.675em%
               {\hrule width.6em\vfil\hrule}\vrule\hfil}}
     \par\vskip3pt}

\newtheorem{theorem}{Theorem}

\input amssym.def

\begin{document}
\title{Method for measuring the entanglement of formation for
    arbitrary-dimensional pure states}
\author{Ming Li$^{1}$}
\author{Shao-Ming Fei$^{2,3}$}
\affiliation{ $^{1}$Department of Mathematics, School of Science,
China University of Petroleum, Qingdao 266555,  China\\
$^2$School of Mathematical Sciences, Capital Normal University,
Beijing
100048, China\\
$^3$Max-Planck-Institute for Mathematics in the Sciences, Leipzig
04103, Germany}

\begin{abstract}
Entanglement of formation is an important measure of quantum entanglement.
We present an experimental way to measure the entanglement of formation
for arbitrary dimensional pure states. The measurement only evolves local
quantum mechanical observables.
\end{abstract}

\maketitle

Quantum entangled states have become the most important physical
resources in quantum communication, information processing and
quantum computing. One of the most difficult and
fundamental problems in entanglement theory is to quantify
the quantum entanglement. A number of entanglement measures such as the
entanglement of formation and distillation \cite{eofd,wootters},
negativity \cite{nega} and concurrence \cite{concu} have been
proposed. Among these entanglement measures, the entanglement of
formation, which quantifies the required minimally physical
resources to prepare a quantum state, plays important roles in
quantum phase transition for various interacting quantum many-body
systems \cite{meof1} and may significantly affect macroscopic
properties of solids \cite{meof2}. Thus the quantitative evaluation of
entanglement of formation is of great significance.

Comparing with the concurrence, entanglement of formation is more
difficult to deal with, and less results have been derived.
The experimental measurement of concurrence has been proposed in \cite{buchleitner}
for pure two-qubit systems by using two copies of the unknown quantum state. In \cite{fei}
the authors presented an approach of measuring concurrence for arbitrary dimensional
pure multipartite systems in terms of only one copy of the unknown quantum state.
However, due to the complicated expression of entanglement of formation,
there is no experimental way yet to determine the entanglement of formation for an unknown
quantum pure state, except for the case of two-qubit systems for which the concurrence and
entanglement of formation have a simple monotonic relations \cite{wootters}.

In this brief report we present an experimental determination of the entanglement of formation
for arbitrary dimensional pure quantum states. The measurement only evolves local
quantum mechanical observables and the entanglement of formation can be obtained according to
the mean values of these observables.

The entanglement of formation is defined for bipartite systems.
Let ${\cal H}_A$ and ${\cal H}_B$ be $m$ and $n$ ($m\leq n$) dimensional complex
Hilbert spaces with orthonormal basis $|i\ra$, $i=1,...,m$, and
$|j\ra$, $i=1,...,n$ respectively. A pure quantum state on ${\cal H}_A \otimes{\cal
H}_B$ is generally of the form,
\begin{equation}\label{psi}
\vert\psi\rangle=\sum_{i=1}^{m}\sum_{j=1}^{n} a_{ij}|ij\ra,~~~~~~a_{ij}\in\Cb
\end{equation}
with normalization
\begin{equation}\label{norm}
\sum_{i=1}^{m}\sum_{j=1}^{n} a_{ij}a_{ij}^\ast=1\,.
\end{equation}
The entanglement of formation of $\vert\psi\rangle$ is defined as the partial entropy with respect to the
subsystems \cite{eofd},
\be\label{epsi}
E(|\psi \rangle) = - {\mbox{Tr}\,} (\rho^A \log_2
\rho^A) = - {\mbox{Tr}\,} (\rho^B \log_2 \rho^B)\,,
\ee
where $\rho^A$ (resp. $\rho^B$) is the reduced density matrix obtained
by tracing  $\bf
|\psi\rangle\langle\psi|$ over the space ${\cal H}_B$ (resp. ${\cal H}_A$).
This definition can be extended to
mixed states $\rho$ by the convex roof,
\begin{equation}\label{erho}
E(\rho )\equiv \min_{\{p_{i},|\psi _{i}\rangle
\}}\sum_{i}p_{i}E({\left\vert \psi _{i}\right\rangle }),
\end{equation}
where the minimization goes over all possible ensemble realizations of $\rho$,
\be\label{rho}
\rho=\sum_{i}p_{i}|\psi _{i}\rangle \langle \psi _{i}|, ~~p_{i}\geq 0,
~~\sum_{i}p_{i}=1.
\ee

A bipartite quantum state $|\psi\ra$ can be
written in the Schmidt form $|\psi\ra=\sum\limits_{i=1}^m
\sqrt{\lambda_i}\,|i_{A}\ra |i_{B}\ra$, $\lambda_{i}\geq 0$,
$\sum_i \lambda_i=1$, under suitable basis $|i_A\ra\in {\cal H}_A$
and $|i_B\ra\in {\cal H}_B$.
$\lambda_i$, $i=1,...,m$, are also the eigenvalues of $\rho^A$.
$E(|\psi\ra)$ can be further expressed as
\be\label{epsilambda}
E(|\psi\ra)=S(\rho^A)=-\sum_{i=1}^m\lambda_i \log\lambda_i.
\ee

For two-qubit case, $m=n=2$, $|\psi\rangle = a_{11}|00\rangle +
a_{12}|01\rangle + a_{21}|10\rangle + a_{22}|11\rangle$, $\vert
a_{11}\vert^2+ \vert a_{12}\vert^2+\vert a_{21}\vert^2+\vert
a_{22}\vert^2=1$. (\ref{epsi}) can be written as \cite{wootters},
\be\label{wootter} E(|\psi\ra)=h(\frac{1+\sqrt{1-C^2}}{2}), \ee
where $h(x)=-x\log_2 x-(1-x)\log_2 (1-x)$,
$C=2|a_{11}a_{22}-a_{12}a_{21}\vert$ is the concurrence. In this
special case $E(|\psi\ra)$ is just a monotonically increasing
function of the concurrence $C$. However for $m \geq 3$, there is no
such relations like (\ref{wootter}) between the entanglement of
formation and concurrence in general. Since for the case $m=2$, due to the normalization condition, $\lambda_1+\lambda_2=1$, only one free parameter is left in the formula (\ref{epsilambda}).
For general high dimensional case, $E(|\psi\ra)$ depends on more free parameters.
Nevertheless, if $\rho^A$ has only two non-zero eigenvalues
(each of which may be degenerate), the maximal
non-zero diagonal determinant $D$ of $\rho^A$ is a generalized
concurrence, namely, the corresponding entanglement of formation
is again a monotonically increasing function of $D$ \cite{fljw}.
The construction of such kind of states is presented in \cite{fei-li}.
In \cite{zhui}, the results are generalized to more general case:
relations like (\ref{wootter}) holds for states with $\rho^A$
having more non-zero eigenvalues such that all these eigenvalues
are functions of two independent parameters.

To measure the quantity (\ref{epsilambda}) experimentally,
we first rewrite the expression (\ref{epsilambda}) according to
the entanglement of formation of some ``two-qubit" states.
Let $L_\alpha^A$ and $L_\beta^B$ be the generators of
special unitary groups $SO(m)$ and $SO(n)$, with the
$m(m-1)/2$ generators $L_\alpha^A$ given by $\{|i\rangle\langle
j|-|j\rangle\langle i|\}$, $1\leq i < j \leq m$, and
the $n(n-1)/2$ generators $L_\beta^B$ given by $\{|k\rangle\langle
l|-|l\rangle\langle k|\}$, $1\leq k < l \leq n$, respectively.
The matrix operators $L_{\alpha}^A$
(resp. $L_{\beta}^B$) have $m-2$ (resp. $n-2$) rows and $m-2$ (resp.
$n-2$) columns that are identically zero.

Let $\rho=|\psi\ra\la\psi|$ be the density matrix with respect to the pure state $|\psi\ra$.
We define
\begin{eqnarray}\label{qus}
\rho_{\alpha\beta}=\frac{L_{\alpha}^A\otimes
L_{\beta}^B\,\rho\,(L_{\alpha}^A)^{\dag}\otimes
(L_{\beta}^B)^{\dag}}{||L_{\alpha}^A\otimes
L_{\beta}^B\,\rho\,(L_{\alpha}^A)^{\dag}\otimes (L_{\beta}^B)^{\dag}||},
\end{eqnarray}
where $\alpha=1,2,\cdots, \frac{m(m-1)}{2}; \beta=1,2,\cdots,
\frac{n(n-1)}{2}$, and $||X||=\sqrt{Tr(XX^{\dag})}$ is the trace norm of matrix $X$.
As the matrix $L_{\alpha}^A\otimes L_{\beta}^B$ has $mn-4$ rows and
$mn-4$ columns that are identically zero, $\rho_{\alpha\beta}$ has
at most $4 \times 4 = 16$ nonzero elements and is called
``two-qubit" state. $\rho_{\alpha\beta}$ is still a normalized pure state.

\begin{theorem}\label{thm}
For any $m\otimes n$ $(m\leq n)$ pure quantum state $|\psi\ra\in {\cal H}_A \otimes{\cal H}_B$,
\be\label{thm}
E(|\psi\ra)=\frac{1}{(m-1)^2}\sum_{\alpha\beta}
\frac{E(\rho_{\alpha\beta})+\log(C_{\alpha\beta})}{C_{\alpha\beta}},
\ee
where $C_{\alpha\beta}=1/Tr\{L_{\alpha}^A\otimes
L_{\beta}^B|\psi\ra\la\psi|(L_{\alpha}^A)^{\dag}\otimes
(L_{\beta}^B)^{\dag}\}$.
\end{theorem}

Proof. To calculate $E(\rho_{\alpha\beta})$ we denote
$L_{\alpha}^A=|a\ra\la b|-|b\ra\la a|$ and $L_{\beta}^B=|c\ra\la
d|-|d\ra\la c|$ for convenience, where $1\leq a<b\leq m$ and $1\leq
c<d\leq m$. Set
\be
\rho^{'}_{\alpha\beta}=L_{\alpha}^A\otimes
L_{\beta}^B|\psi\ra\la\psi|(L_{\alpha}^A)^{\dag}\otimes
(L_{\beta}^B)^{\dag}.
\ee
It is direct to verify that
\be
\rho^{'}_{\alpha\beta}=|\psi\ra_{\alpha\beta}\la\psi|,
\ee
where $|\psi\ra_{\alpha\beta}=\lambda_b \delta_{bd}|ac\ra-\lambda_b
\delta_{bc}|ad\ra-\lambda_a \delta_{ad}|bc\ra+\lambda_a
\delta_{ac}|bd\ra$.

We now compute the eigenvalues of $\rho_{\alpha\beta}^{'A}=Tr_B(\rho_{\alpha\beta}^{'})$
according to the values of $a, b, c$ and $d$:

i). $a\neq b\neq c\neq d$. We have $|\psi\ra_{\alpha\beta}=0$.

ii). $b>a=c<d$ and $b\neq d$. We get
$|\psi\ra_{\alpha\beta}=\sqrt{\lambda_a}|bd\ra$. The eigenvalue of
$\rho_{\alpha\beta}^{'A}$ corresponding to this case is
$\lambda_a$. As $a=c$ can be chosen to be $1, 2,
\cdots, m-2$, $b$ and $d$ have only $m-k$ and $m-k-1$
(corresponding to $a=c=k$, $k=1, 2, \cdots, m-2$) kinds of choices. Altogether
we have $(m-k)(m-k-1)$ eigenvalues of
$\rho_{\alpha\beta}^{'A}$ to be $\lambda_k$ in this case, with
$k=1, 2, \cdots, m-2$.

iii). $a<b=d>c$ and $a\neq c$. We have
$|\psi\ra_{\alpha\beta}=\sqrt{\lambda_b}|ac\ra$. The eigenvalue of
$\rho_{\alpha\beta}^{'A}$ is $\lambda_b$. In this case $b=d$ can be $3, 4,
\cdots, m$. Then $a$ and $c$ have only $k-1$ and $k-2$
(corresponding to $b=d=k$, $k=3, 4, \cdots, m$) kinds of choices. Hence we have $(k-1)(k-2)$ eigenvalues of
$\rho_{\alpha\beta}^{'A}$ to be $\lambda_k$ in this case, $k=3, 4, \cdots, m$.

iv). $b>a=c<d$ and $b=d$. We obtain $|\psi\ra_{\alpha\beta}=\sqrt{\lambda_b}|ac\ra+\sqrt{\lambda_a}|bd\ra$.
The eigenvalues of $\rho_{\alpha\beta}^{'A}$ are $\lambda_a$ and $\lambda_b$, and $a=c$
can be $1, 2, \cdots, m-1$. Then $b=d$ can be $k+1, k+2, \cdots, m$ (corresponding to $a=c=k, k=1, 2, \cdots,
m-1$). We have $(m-1)$ eigenvalues of
$\rho_{\alpha\beta}^{'A}$ to be $\lambda_k$, $k=1, 2, \cdots, m$.

v). $a<b=c<d$. We have
$|\psi\ra_{\alpha\beta}=-\sqrt{\lambda_b}|ad\ra$. The eigenvalue of
$\rho_{\alpha\beta}^{'A}$ is $\lambda_b$. $b=c$ can be $2, 3,
\cdots, m-1$. Then $a$ and $d$ have only $k-1$ and $m-k$
(corresponding to $b=c=k$, $k=2, 3, \cdots, m-1$) kinds of choices.
We have $(k-1)(m-k)$ eigenvalues of
$\rho_{\alpha\beta}^{'A}$ to be $\lambda_k$, $k=2, 3, \cdots, m-1$.

vi). $c<d=a<b$. We have
$|\psi\ra_{\alpha\beta}=-\sqrt{\lambda_a}|bc\ra$. The eigenvalue of
$\rho_{\alpha\beta}^{'A}$ is
$\lambda_a$. In this case $a=d$ can be $2, 3,
\cdots, m-1$. $c$ and $b$ have only $k-1$ and $m-k$
(corresponding to $b=c=k$, $k=2, 3, \cdots, m-1$) kinds of choices.
Therefore we have $(k-1)(m-k)$ eigenvalues of
$\rho_{\alpha\beta}^{'A}$ to be $\lambda_k$, with $k=2, 3, \cdots, m-1$.

Let $\lambda^i_{\alpha\beta}$ stand for the eigenvalues of
$\rho_{\alpha\beta}^A$. From the analysis of cases i)-vi) and
formula (\ref{epsilambda}), we get
\begin{eqnarray}\label{bop}
E(|\psi\ra)=-\frac{1}{(m-1)^2}\sum_{\alpha\beta}\sum_{i=1}\lambda_{\alpha\beta}^i\log(\lambda_{\alpha\beta}^i).
\end{eqnarray}

Since
\be
\rho_{\alpha\beta}=\frac{\rho^{'}_{\alpha\beta}}{Tr\{\rho^{'}_{\alpha\beta}\}}=C_{\alpha\beta}\rho^{'}_{\alpha\beta},
\ee
we have
$\sum_i\lambda_{\alpha\beta}^iC_{\alpha\beta}=1$ for any $\alpha$ and $\beta$.
Therefore
\begin{eqnarray}
E(\rho_{\alpha\beta})&&=-\sum_{i=1}C_{\alpha\beta}\lambda_{\alpha\beta}^i
\log(C_{\alpha\beta}\lambda_{\alpha\beta}^i)\nonumber\\
&&=-\sum_{i=1}C_{\alpha\beta}\lambda_{\alpha\beta}^i\log(C_{\alpha\beta})
-\sum_{i=1}C_{\alpha\beta}\lambda_{\alpha\beta}^i\log(\lambda_{\alpha\beta}^i)\nonumber\\
&&=-\log(C_{\alpha\beta})-C_{\alpha\beta}\sum_{i=1}\lambda_{\alpha\beta}^i\log(\lambda_{\alpha\beta}^i).\nonumber
\end{eqnarray}
That is
\be\label{pr}-\sum_{i=1}^{m}\lambda_{\alpha\beta}^i\log(\lambda_{\alpha\beta}^i)
=\frac{E(\rho_{\alpha\beta})+\log(C_{\alpha\beta})}{C_{\alpha\beta}}.\ee
Substituting (\ref{pr}) into (\ref{bop}), we obtain that
\be
E(|\psi\ra)=\frac{1}{(m-1)^2}\sum_{\alpha\beta}\frac{E(\rho_{\alpha\beta})+\log(C_{\alpha\beta})}{C_{\alpha\beta}},
\ee
which proves the theorem. \qed

The theorem shows that one can derive the entanglement of formation of a pure
quantum state by measuring the values of the entanglement of formation of
all the states $\rho_{\alpha\beta}$ and values of
$C_{\alpha\beta}$. Here if $|\psi\ra_{\alpha\beta}=0$, then
$C_{\alpha\beta}$ goes to infinity and this term does not
contribute to the summation in (\ref{thm}).
Hence the summation $\sum_{\alpha\beta}$ in (\ref{thm}) simply
goes over all the terms such that $Tr\{L_{\alpha}^A\otimes
L_{\beta}^B|\psi\ra\la\psi|(L_{\alpha}^A)^{\dag}\otimes
(L_{\beta}^B)^{\dag}\}\neq 0$.

With formula (\ref{thm}), we now show how to get the value of $E(|\psi\ra)$ experimentally by measuring
the quantities on the right hand side of (\ref{thm}).

The quantity $C_{\alpha\beta}={1}/{Tr\{\rho^{'}_{\alpha\beta}\}}$
can be determined by $Tr\{\rho^{'}_{\alpha\beta}\}$. Since
$Tr\{\rho^{'}_{\alpha\beta}\}=\la\psi|(L_{\alpha}^A)^{\dag}L_{\alpha}^A\otimes
(L_{\beta}^B)^{\dag}L_{\beta}^B|\psi\ra$, one can obtain
$C_{\alpha\beta}$ by measuring the
local Hermitian operator $(L_{\alpha}^A)^{\dag}L_{\alpha}^A\otimes
(L_{\beta}^B)^{\dag}L_{\beta}^B$ associated with the state $|\psi\ra$.

To measure $E(\rho_{\alpha\beta})$, we first note that although
$\rho_{\alpha\beta}$ are $m\otimes n$ bipartite quantum states,
they are basically ``two-qubit" ones. For given $L_{\alpha}=|i\ra\la j|-|j\ra\la i|$ and
$L_{\beta}=|k\ra\la l|-|l\ra\la k|$, $i\neq j$, $k\neq l$, the non-zero elements
of $\rho_{\alpha\beta}$ are located at the $i*(m-1)+k$th, $i*(m-1)+l$th,
$j*(m-1)+k$th, and $j*(m-1)+l$th rows and the $i*(m-1)+k$th,
$i*(m-1)+l$th, $j*(m-1)+k$th, and $j*(m-1)+l$th columns. They constitute a $4\times 4$ matrix,
$$
\sigma_{\alpha\beta}^{'}=\left(
    \begin{array}{ccccccccc}
      \rho_{ik,ik} & \rho_{ik,il} & \rho_{ik,jk} & \rho_{ik,jl}\\
      \rho_{il,ik} & \rho_{il,il} & \rho_{il,jk} & \rho_{il,jl}\\
      \rho_{jk,ik} & \rho_{jk,il} & \rho_{jk,jk} & \rho_{jk,jl}\\
      \rho_{jl,ik} & \rho_{jl,il} & \rho_{jl,jk} & \rho_{jl,jl}
    \end{array}\right).
$$
Set $\sigma_{\alpha\beta}={\sigma^{'}_{\alpha\beta}}/{Tr\{\sigma^{'}_{\alpha\beta}\}}$.
Obviously $E(\rho_{\alpha\beta})=E(\sigma_{\alpha\beta})$.
But $\sigma_{\alpha\beta}$ are actually two-qubit pure states. According to
the formula (\ref{wootter}), $E(\sigma_{\alpha\beta})$ is
determined by the concurrence $C(\sigma_{\alpha\beta})=C(\rho_{\alpha\beta})$.
Therefore if we can measure the quantity $C(\rho_{\alpha\beta})$, we can obtain $E(\rho_{\alpha\beta})$.

The quantity $C(\rho_{\alpha\beta})$ can be measured experimentally in
terms of the method introduced in \cite{fei}, with a few
modifications of the measurement operators. Corresponding to the case of
$L_{\alpha}=|i\ra\la j|-|j\ra\la i|$ and $L_{\beta}=|k\ra\la l|-|l\ra\la k|$, we define
$m\times m$ matrix operators $\Sigma_s$, $s=0,1,2,3$, such that
$(\Sigma_0)_{pq}=\delta_{pi}\delta_{qi}+\delta_{pj}\delta_{qj}$,
$(\Sigma_1)_{pq}=\delta_{pi}\delta_{qj}+\delta_{pj}\delta_{qi}$,
$(\Sigma_2)_{pq}=I \delta_{pi}\delta_{qj}-I\delta_{pj}\delta_{qi}$,
$(\Sigma_3)_{pq}=\delta_{pi}\delta_{qi}-\delta_{pj}\delta_{qj}$, $p,q=1,...,m$.
Similarly we define $n\times n$ matrix operators $\Gamma_0, \Gamma_1,
\Gamma_2$ and $\Gamma_3$ by replacing the indices $i,j$ in
$\Sigma_0, \Sigma_1, \Sigma_2$ and $\Sigma_3$ with $k,l$ respectively, and setting $p,q=1,...,n$.
It is straightforward to derive that
$C(\rho_{\alpha\beta})$ can be expressed as the mean values of the above local observables,
\begin{eqnarray}
C^2(\rho_{\alpha\beta})&=&\frac{1}{2}+\frac{C^2_{\alpha\beta}}{2}\left(\la\Sigma_3\otimes\Gamma_3\ra^2
-\la\Sigma_3\otimes\Gamma_0\ra^2\right.\nonumber\\
&&-\la\Sigma_0\otimes\Gamma_3\ra^2-\la\Sigma_0\otimes\Gamma_1\ra^2+\la\Sigma_3\otimes\Gamma_1\ra^2\nonumber\\
&&\left.-\la\Sigma_0\otimes\Gamma_2\ra^2+\la\Sigma_3\otimes\Gamma_2\ra^2\right).
\end{eqnarray}

In summary we have presented an experimental way to measure the entanglement of formation
for arbitrary dimensional pure states, by measuring some local
quantum mechanical observables. We reduced the difficult problem to find the
concurrence of ``two-qubit" states for which many results have been already derived.
Recently high dimensional bipartite systems like in NMR and nitrogen-vacancy defect center
have been successfully used for quantum computation and simulation experiments \cite{du}.
Our results present a plausible way to measure the entanglement of formation in these systems and
to investigate the roles played by the entanglement of formation in these quantum information processing.

So far experimental measurement on entanglement of formation and concurrence concerns only pure states.
For mixed states, less is known except for experimental determination of separability, both sufficiently
and necessary, for two-qubit \cite{yusx} and qubit-qutrit systems \cite{zhmj}.
Generally (\ref{erho}) has only analytical results for some special states \cite{iso} and
analytical lower bounds \cite{elb} which are not experimentally measurable.
Recently in \cite{emlb} we have presented a measurable lower bound of entanglement of formation.
The formula (\ref{thm}) for pure state may also help to study measurable lower bounds of entanglement of formation
for mixed states.

\bigskip
\noindent{\bf Acknowledgments}\, This work is supported by the NSFC 10875081 and PHR201007107.

\end{document}